\documentstyle[twocolumn,aps]{revtex}

\begin{document}
\draft
\preprint{hep-ph/9809508\\}
\title{Another Look at Just-So Solar Neutrino Oscillations\\
}
\author{James M. Gelb\cite{gelb}\\}
\address{
Department of Physics\\
University of Texas at Arlington\\
Arlington, Texas 76019\\
}
\author{S. P. Rosen\cite{rosen}\\}
\address{
U.S. Department of Energy\\
Germantown, Maryland 20874\\
}
\date{\today}
\maketitle

\begin{abstract}
We take another look at ``Just-So'' solar neutrino oscillations,
characterizing them by the energy
$E_{\pi/2}$ at which the distance-varying angle is $\pi/2$, 
instead of by the
usual $\Delta m^2$. The rising
spectrum recently observed by SuperKamiokande is consistent with an
$E_{\pi/2}$ $\sim 6-9$ MeV and marginally with 48 MeV. The
pp neutrinos must then be reduced to one-half the standard solar model
prediction, and $^7$Be
neutrinos must make up a significant part of the SAGE and GALLEX
gallium signal. For $E_{\pi/2}$
close to 9 and 48 MeV, the $^7$Be neutrinos will also show a large seasonal
variation, emphasizing the
importance of direct
measurements of the $^7$Be neutrinos.
\end{abstract}

\narrowtext

According to recent results from SuperKamiokande\cite{superk}, there is an
excess of events at the
high-energy end of the solar neutrino spectrum relative to the standard
solar model (SSM)\cite{ssm}
and the small-angle MSW solution of the solar neutrino
problem\cite{smallang}.
In particular, the last
three points above 12 MeV appear to rise more rapidly than a simple
comparison with the SSM
and MSW would indicate. This excess suggests that solar neutrinos may
be undergoing ``Just-So''
{\it in vacuo} oscillations with large mixing\cite{gkk98,justso} instead of
matter-enhanced MSW oscillations with
small mixing, and that the relevant $\Delta m^2$ is of order
$10^{-10}~{\rm eV}^2$ rather than $10^{-5}~{\rm eV}^2$.

The key point about ``Just-So''
oscillations\cite{justso} is that the neutrino
oscillation length is
sufficiently close to the distance between the Sun and the Earth that
the oscillation probability for
the higher energy $^8$B solar neutrinos is in the region of slow
oscillations rather than the rapid
oscillations at the lower end of the solar neutrino spectrum. This
means that instead of taking on
the average value of one-half, the survival probability of $^8$B
neutrinos depends upon their
energy and can run the gamut from zero to one. It thus becomes possible
for the ``Just-So''
electron-neutrino survival probability to exceed that of the small
mixing-angle MSW case above some
specific energy\cite{app}; and this effect can then yield the qualitative
behavior observed by SuperKamiokande.

To mark the division between slow and fast oscillations in terms of
neutrino energy, it is
convenient to characterize the oscillations by the energy at which the
distance-dependent angle is $\pi/2$:
\begin{equation}
1.27~\Delta m^2~{d_\odot\over E_{\pi/2}}~=~{\pi\over 2}~;~~
\Delta m^2 = 8.27\times 10^{-12}~E_{\pi/2}~.	
\end{equation}
For a fixed solar distance $d_\odot$, $1.5\times 10^8$ km 
chosen here, this is entirely equivalent
to the usual $\Delta m^2$
characterization with energies in MeV and $\Delta m^2$ in eV$^2$. 
At $E_{\pi/2}$, the survival probability of solar
electron-type neutrinos as they
travel to Earth from the Sun:
\begin{equation}
P(\nu_e \rightarrow\nu_e;~E_\nu) = 1 - \sin^2 2\theta
\sin^2\left({\pi\over 2}{E_{\pi/2}\over E_\nu}\right)
\end{equation}
reaches its minimum value of $(1- \sin^2 2\theta)$ and then increases
monotonically to 1 as $E_\nu$
increases.

By choosing $E_{\pi/2}$ to be sufficiently below the endpoint energy of
the $^8$B neutrinos, we can
ensure that the survival probability for ``Just-So'' oscillations exceeds
0.4, the approximate value
predicted by the small-angle MSW solution at the high-energy end of the
solar neutrino
spectrum\cite{smallang}. This will give rise to a ratio of the ``Just-So''
spectrum to the SSM 
spectrum which turns up more rapidly at the high-energy end than
the corresponding ratio
for the MSW solution.

For neutrino energies smaller than $E_{\pi/2}$, oscillations set in and
become more rapid as $E_{\nu}$
decreases. The survival probability reaches its maximal value at
$E_{\pi/2}/2$ and its minimum value
at $E_{\pi/2}/3$; the pattern then continues 
with $P(\nu_e \rightarrow\nu_e;~E_\nu)$
maximal for $E_\nu$ an even
divisor of $E_{\pi/2}$ and becoming minimal for $E_\nu$
an odd divisor. The
smaller the energy, the
more rapidly the oscillations take place. This behavior is illustrated
in Fig.~1.

In Fig.~2 we plot the ratio of differential cross-sections for $^8$B
neutrinos
for several choices of $E_{\pi/2}$ in the
case of maximal mixing $(\sin^2
2\theta = 1)$ to the no-oscillation scenario. 
We use the standard cross-sections\cite{standcross} and take
into account the smearing
effect of the energy resolution\cite{standcross,smear}. 
For the choice of 6 MeV the
curve begins to turn up at
4.5 MeV, and for the choice of 9 MeV it begins to turn up at
8.5 MeV.
For
larger values of
$E_{\pi/2}$, for example 12 and 15 MeV, the curves fail to increase at
higher energy.
However, for even higher $E_{\pi/2}$, for example 48 MeV,
which corresponds to the best allowed
parameter for
``Just-So'' oscillations\cite{phenom}, 
the curve turns up 
at 8 MeV but begins to turn over at 13.5 MeV.
Of the $\sin^22\theta=1$ cases considered here
that turn up at the high-energy end,
the $E_{\pi/2} =$ 9 and 48 MeV cases are acceptable 
parameters for ``Just-So'' oscillations,
with the 6 MeV case marginally allowed\cite{phenom}.

It is not difficult to see that for the above $E_{\pi/2}$, both
the pp and the $^7$Be
neutrinos, being of much lower energy, are in the region of rapid
oscillations. For the pp
neutrinos, this implies that their $\nu_e$ survival probability will
reach the average value of
\begin{equation}
<P(\nu_e \rightarrow \nu_e;~E_\nu)> = 1 - {1\over 2}\sin^2 2\theta~.
\end{equation}
For $^7$Be neutrinos, the value of the survival probability is a more
delicate matter since they are
monoenergetic; it is given precisely by
\begin{equation}
P(\nu_e \rightarrow\nu_e;0.86{\rm MeV}) = 1 - \sin^2 2\theta
\sin^2\left({\pi\over 2}{E_{\pi/2}\over 0.86}\right)~,
\end{equation}
and will vary between its minimal and maximal values depending on the
precise value of $E_{\pi/2}$.
(It should be noted 
that variations in solar distances associated with the size
of the $^7$Be production
zone $(\sim 10^5~{\rm km})$ are 
too small compared to the mean Earth-Sun distance
$(\sim 10^8~{\rm km})$ to affect
the survival probability by more than 1-2\%.)

In order to estimate the survival probability for $^7$Be neutrinos in
this picture, we appeal to the
results of the SAGE and GALLEX experiments\cite{sagegallex} 
which indicate that
the total signal is
almost exactly equal to the signal predicted by the SSM
for pp neutrinos
alone\cite{ssm}. Since the pp neutrinos in this ``Just-So'' case are reduced
to one-half their SSM
value, the $^7$Be neutrinos must help to make up the difference. The
SSM prediction for $^7$Be
neutrinos is approximately one-half the prediction for pp
neutrinos\cite{ssm} and so a large fraction
of these, possibly all, will be needed to make up for the reduced
signal from the pp neutrinos
themselves. To meet this requirement the value of $E_{\pi/2}$ must be
``fine-tuned'' so that its ratio to
the $^7$Be energy, 0.86 MeV, is close to an odd number---another
``Just-So'' condition!

This prediction for the survival probability of $^7$Be neutrinos
should be contrasted with that of
the small-angle MSW solution, which requires that they must be largely
converted to non-electron
neutrinos\cite{gkr92}. It should also be considered in light of
phenomenological analyses\cite{phenom}
of the chlorine plus SuperKamiokande experiments, which
indicates the possibility (but not the requirement) that the
the $^7$Be contribution may not dominate the gallium signal. This 
would tend to
disfavor the ``Just-So''
explanation
of the spectral effect, and to lend support to the small-angle MSW
solution plus an unexpectedly
large contribution from hep neutrinos at the high-energy
end\cite{kb_hep}.

Another more subtle consequence of the ``Just-So'' explanation is that,
as has been pointed out by
many authors\cite{gkk98,many_season},
the survival 
probability for $^7$Be neutrinos
may have a significant
seasonal variation due to the eccentricity of the orbit of the
Earth. The variation in the
orbit of the Earth due to its eccentricity is of order $10^6$ km, and
this can induce a large change in
the distance-dependent angle, see Eq.~1, if 
the ratio $E_{\pi/2}/0.86$
happens to be large. Should
this be the case, then gallium experiments like SAGE and 
GALLEX\cite{sagegallex}
will also show a
seasonal variation.

In terms of the
eccentricity of the Earth's orbit
around the Sun,
$\epsilon = 0.017$, 
the oscillating factor in the probability is
then given by:
\begin{equation}
P(\nu_e \rightarrow\nu_e) = 1 - \sin^2 2\theta \sin^2 \left[{\pi\over 2}
{E_{\pi/2}\over 0.86}{{1-\epsilon^2}\over 1 + \epsilon \cos \phi}\right]~.
\end{equation}
Below we plot Eq.~5 and we also include curves with the $1/d^2$ 
variation in the neutrino flux
which slightly enhances or diminishes the probability relative to
oscillations alone\cite{gkk98}.
If  $E_{\pi/2}$ is in the neighborhood of 9 MeV, then the angular factor
will be close to $5\pi +
2\pi/3$ and the probability will vary from 0.5 to 1, with a mean
value of 0.75, during the year.
On the other hand, if the factor is closer to 4.5$\pi$, the variation
will only be about 5\%.

In order to calculate the impact of this eccentricity dependence on
gallium experiments, we need
to choose the mixing angle. For the sake of illustration, we take the
``best value'' of Bahcall,
Krastev, and Smirnov\cite{phenom}:
\begin{equation}
\sin^2 2\theta = 0.75~,
\end{equation}
with probabilities shown in Fig.~3 for three values of $E_{\pi/2}$.
The gallium signal from pp neutrinos is then
\begin{equation}
S_{\rm pp} = 72\times(1- 0.75\times0.5) = 72\times(5/8) = 45~{\rm SNU}~,
\end{equation}
and the yearly mean from $^7$Be neutrinos is
\begin{equation}
S_{^7{\rm Be}} = 35\times (1 - 0.75\times 0.75) = 16~{\rm SNU}~.
\end{equation}
Taking an average signal of 73 SNU, we assume that the remaining 12 SNU
come from a
combination of about 5 SNU from $^8$B and 7 SNU from other solar
neutrinos\cite{ssm}, and that
these neutrinos, which are generally higher in energy than $^7$Be, do
not vary appreciably with
the seasons.  The seasonal variation of $^7$Be neutrinos runs from 22
SNU in winter to 9 SNU in
summer, and so there could be a variation of $\pm 10\%$ in the total
gallium signal\cite{sagegallex}. (The effect is even larger for
the 48 MeV case.)

Experiments designed to detect the $^7$Be alone\cite{borexino}, 
should see the full extent of
the seasonal variation.  Alternatively, one could
analyze the gallium results
under the assumption that the observed signal consists of one-half the
SSM prediction for pp
neutrinos, about 36 SNU, plus about 10 from $^8$B and other neutrinos,
and the remainder from
$^7$Be. One could therefore consider the variation of the quantity
\begin{equation}
Q = {\rm observed~Ga~signal}~-~46~{\rm SNU}
\end{equation}
which will amplify, on a percentage basis, the variation of the $^7$Be
signal in this picture.

Independently of possible seasonal variations,  a key test of these
ideas is provided by the average
$^7$Be neutrino signal. For ``Just-So'' oscillations, $^7$Be 
neutrinos must
significantly remain as
electron neutrinos and the signal will be large; whereas for the
small-angle MSW solution, they
must largely be converted to non-electron types and the signal will be
small. Therefore, as long as
the hints for ``Just-So'' oscillations persist\cite{superk}, experiments
designed to detect $^7$Be
neutrinos will be highly important in confirming, or invalidating them.

\begin{figure}
\caption{
Oscillation survival probability, $\nu_e$ to $\nu_e$, as a function of 
$E_\nu/E_{\pi/2}$, where $E_{\pi/2}$ is
the energy at which the distance-dependent angle equals 
$\pi/2$. It shows the slowly varying region for energies greater than 
$E_{\pi/2}$ and the rapidly varying region below $E_{\pi/2}/3$. 
(Values are not shown below $E_{\pi/2}/8$ for clarity.)
}
\label{fig1}
\end{figure}

\begin{figure}
\caption{
The ratios of differential
cross-sections for $^8$B neutrino spectrum
and SuperKamiokande resolution function
for various values of $E_{\pi/2}$ to the SSM differential 
cross-sections are shown.  
The curves for $E_{\pi/2}$ in the range of 6 to 9 MeV 
(and possibly 48 MeV) appear to correspond to the energy spectrum observed by
SuperKamiokande.
}
\label{fig2}
\end{figure}

\begin{figure}
\caption{
Seasonal variation of the $^7$Be survival probability for 
various values of $E_{\pi/2}$. In particular, the variation 
is large for
$E_{\pi/2}$ = 9 and 48 MeV and small for 6 MeV.
Solid lines are for $P(\nu_e\rightarrow\nu_e;~d)$ alone 
(Eq.~5) and dashed lines
are for $P(\nu_e\rightarrow\nu_e;~d)/d^2$, with $d$ a function of 
orbital phase angle $\phi$.
}
\label{fig3}
\end{figure}

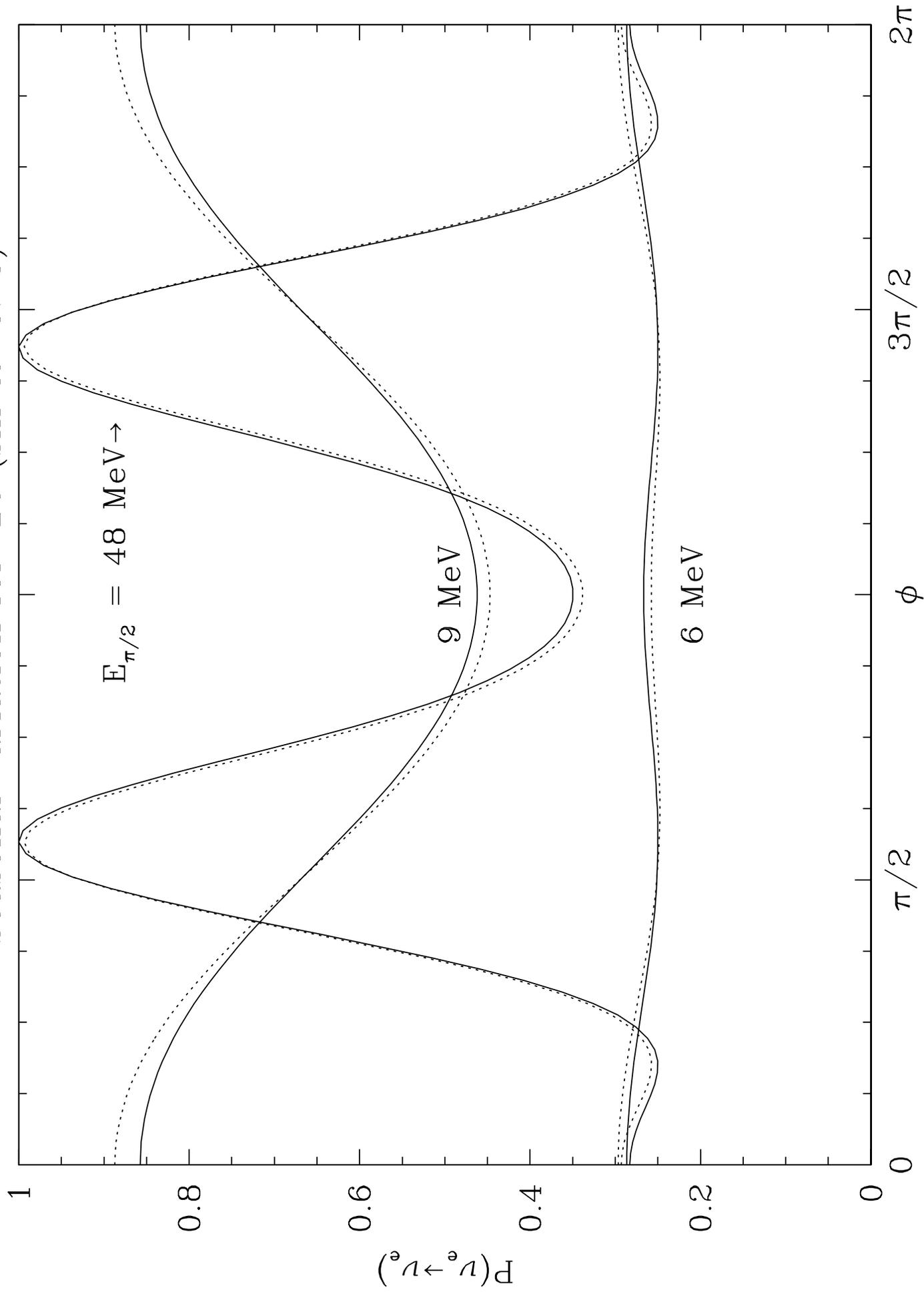
\end{document}